\pdfoutput=1
\documentclass[12pt,a4]{article}
\usepackage{a4wide}
\usepackage[hidelinks]{hyperref}
\usepackage{latexsym}
\usepackage{cite}
\usepackage{mathrsfs}
\usepackage{amssymb}
\usepackage{amsmath}
\usepackage{graphicx}
\usepackage{braket}
\usepackage{enumerate}
\usepackage[usenames,dvipsnames]{xcolor}
\usepackage{todonotes}
\usepackage{breqn}

\newcommand{\cN}{\mathcal{N}}

\newcommand{\be}{\begin{equation}\label}
\newcommand{\ee}{\end{equation}}
\newcommand{\bea}{\begin{eqnarray}\label}
\newcommand{\eea}{\end{eqnarray}}

\begin{document}
 \begin{flushright} DCPT-17/47
 	\end{flushright}
\title{M-theory Beyond The Supergravity Approximation }
\author{Paul Heslop and Arthur E. Lipstein \vspace{7pt}\\ \normalsize \textit{
Department of Mathematical Sciences}\\\normalsize\textit{Durham University, Durham, DH1 3LE, United Kingdom}}

{\let\newpage\relax\maketitle}
\begin{abstract}
We analyze the four-point function of stress-tensor multiplets for the 6d quantum field theory with $OSp(8^*|4)$ symmetry which is conjectured to be dual to M-theory on $AdS_7 \times S^4$, and deduce the leading correction to the tree-level supergravity prediction by obtaining a solution of the crossing equations in the large-$N$ limit with the superconformal partial wave expansion truncated to operators with zero spin. This correction corresponds to the M-theoretic analogue of $\mathcal{R}^4$ corrections in string theory. We also find solutions corresponding to higher-spin truncations, but they are subleading compared to the 1-loop supergravity prediction, which has yet to be calculated.
\end{abstract}
\pagebreak
\tableofcontents

\section{Introduction}

One of the most pressing open questions in string theory is how to formulate M-theory, which arises in the strong-coupling limit of type IIA string theory \cite{Townsend:1995kk,Witten:1995ex}. Although this is a very difficult question, it is known that M-theory reduces to 11d supergravity at low energies \cite{Cremmer:1978km}, and its fundamental degrees of freedom are 2d and 5d objects known as M2-branes and M5-branes, respectively. The worldvolume theory for M2-branes (on an orbifold) was found to be a 3d superconformal Chern-Simons matter theory with classical $\mathcal{N}=6$ supersymmetry \cite{Aharony:2008ug}. On the other hand, the M5-brane worldvolume theory is expected to be a 6d superconformal quantum field theory with $OSp(8^*|4)$ symmetry, and we will subsequently refer to it as the 6d $(2,0)$ theory. For the case of a single M5-brane, it can be formulated in terms of an abelian $(2,0)$ tensor multiplet, which consists of a self-dual two-form gauge field, five scalars, and eight fermions \cite{Howe:1983fr,Perry:1996mk,Pasti:1997gx}, but it is unclear how to generalize this construction to describe multiple M5-branes. A crucial hint is provided by the AdS/CFT conjecture which states that the worldvolume theory for a stack of $N$ M5-branes is dual to M-theory on $AdS_7 \times S^4$ with $N$ units of flux through the 4-sphere, which reduces to 11d supergravity on this background in the limit $N \rightarrow \infty$ \cite{Maldacena:1997re}.     

The goal of this paper is to initiate an approach for extending this description of multiple M5-branes away from $N\rightarrow \infty$, or equivalently beyond the supergravity approximation. Since it is unclear how to formulate the 6d $(2,0)$ theory, our strategy will be to use superconformal and crossing symmetry to deduce the structure of the four-point correlator of stress-tensor multiplets in this theory. Note that the stress tensor belongs to a 1/2-BPS multiplet whose superconformal primary, $T_{IJ}$, is a dimension-4 scalar in the two-index symmetric traceless representation ({\bf 14}) of the R-symmetry group $SO(5) \sim Sp(4)$, whose supergravity dual is a scalar in the graviton multiplet with AdS mass $m^2=-8$ (in units of the inverse AdS radius). This is closely related to a broader strategy known as the conformal bootstrap program, which was pioneered long ago in \cite{Ferrara:1973yt,Ferrara:1973vz,Polyakov:1974gs} and revived more recently following \cite{Rattazzi:2008pe}. More concretely, the idea of the conformal bootstrap is to use the operator product expansion (OPE) to decompose four-point correlators of primary operators into a sum over intermediate operators labelled by their spin and scaling dimension. The contribution of an intermediate primary operator and its descendants can be encoded in a function of two conformal cross-ratios known as a conformal partial wave, and the coefficients of the OPE expansion encode the three-point functions of primary operators. Using crossing symmetry to equate the OPE expansion of a four-point correlator in two different channels then implies powerful constraints on the scaling dimensions and OPE coefficients (collectively known as the CFT data). Note that the existence of local operators in the 6d $(2,0)$ theory is implied by the AdS/CFT conjecture for $N\rightarrow \infty$ and is certainly also the case for $N=1$, so we will assume that local operators exist for any $N$. 

Our work on stress-tensor correlators in the 6d $(2,0)$ theory builds upon several previous papers~\cite{Howe:1983fr,hep-th/9812133,hep-th/9807186,hep-th/9902153,Bastianelli:1999ab,Bastianelli:1999vm,hep-th/0008048,Eden:2001wg,Arutyunov:2002ff,Dolan:2003hv,Dolan:2004mu,Heslop:2004du}. For example, 3-point correlators were studied in \cite{Bastianelli:1999ab,Bastianelli:1999vm,Eden:2001wg} and four-point correlators were derived  in the supergravity approximation in \cite{Arutyunov:2002ff}. A superconformal partial wave (CPW) analysis of these results was subsequently carried out in \cite{Heslop:2004du} using a supersymmetric generalization of the conformal partial waves derived in \cite{Dolan:2003hv} (see also~\cite{Dolan:2004mu}). In \cite{Beem:2015aoa}, the numerical bootstrap was used to obtain bounds on CFT data which are saturated by supergravity predictions in the $N\rightarrow \infty$ limit. The approach we take in this paper is inspired by \cite{Heemskerk:2009pn}, which devised a method to analytically compute CFT data in a generic 2d or 4d CFT by solving the crossing equations for 4-point functions in the large-$N$ limit and truncating the CPW expansion to finite spin\footnote{For a CFT dual to string theory, the number of degrees of freedom scales like $N^2$ in the large-$N$ limit, but for the 6d $(2,0)$ theory it should scale like $N^3$ \cite{Klebanov:1996un}. Hence, the 6d crossing equations should be expanded in $1/N^3$.}.  Moreover, they showed that the solutions to the crossing equations are in one-to-one correspondence with local quartic interaction vertices for a massive scalar field in AdS (which can be thought of as a toy model for the low-energy effective theory in the bulk), and that the large-twist behaviour of the anomalous dimensions derived from these solutions is related to the number of derivatives appearing in the bulk interaction vertices. 

These methods were adapted to $\mathcal{N}=4$ super-Yang-Mills (SYM)  in \cite{Alday:2014tsa} and in this paper we adapt them to the 6d $(2,0)$ theory. In particular, we construct analytical solutions to crossing equations truncated to spin-0,2, and 4 and find the same number of solutions as in 2d and 4d. Although we cannot determine the numerical coefficients of these solutions, we can apply the arguments of \cite{Heemskerk:2009pn} to deduce how these solutions scale in the large-$N$ limit. We find that the spin-0 solution scales like $N^{-5}$, whereas all the higher spin solutions are suppressed by at least $N^{-19/3}$. The tree-level supergravity prediction for the 4-point function scales like $N^{-3}$ and so the one-loop supergravity prediction (which has yet to be computed) should scale like $N^{-6}$, this implies that the spin-0 solution is the {\em leading} correction to the supergravity prediction for the four-point function of stress-tensor correlators and that higher-spin solutions are subleading compared to the finite part of the 1-loop supergravity prediction. The analogous spin-0 solution to the crossing equations in $\mathcal{N}=4$ SYM, obtained in \cite{Alday:2014tsa}, corresponds schematically to an $\mathcal{R}^4$ correction to 10d supergravity in $AdS_5 \times S^5$ (where $\mathcal{R}$ is the Riemann tensor), and can therefore be derived using perturbative string theory (see \cite{Sakai:1986bi,Grisaru:1986dk,Gross:1986iv} for the derivation of such terms in flat background). On the other hand, the spin-0 solution we obtain corresponds to an M-theoretic correction to 11d supergravity in $AdS_7 \times S^4$ and there is currently no other method to derive this from first principles, although such corrections have been deduced in flat background using arguments based on dimensional reduction \cite{Green:1997di,Green:1997as,Russo:1997mk}.

The structure of this paper is as follows. In section \ref{cpw} we review the CPW expansion for the 6d $(2,0)$ theory following \cite{Heslop:2004du}, and in section \ref{sugra} we apply it to the supergravity prediction for the 4-point function of stress-tensor multiplets to deduce the anomalous dimensions and OPE coefficients. In section \ref{corrections}, we obtain corrections to the supergravity prediction for the 4-point function by solving the crossing equations truncated to spin-0,2, and 4. Finally, we present our conclusions in section \ref{conclusion}.  

\section{Superconformal Partial Wave Expansion} \label{cpw}

Superconformal symmetry in six dimensions has been studied in a number of papers~\cite{Howe:1983fr,hep-th/9812133,hep-th/0008048,hep-th/9807186,hep-th/9902153,Eden:2001wg,Arutyunov:2002ff,Dolan:2004mu,Heslop:2004du}.
It constrains the four-point function of stress-tensor multiplets in the 6d $(2,0)$ theory in terms of a single function, or 
``prepotential", of two variables $F(x_1,x_2)$ as follows
\[
\lambda^{4}\left(g_{13}g_{24}\right)^{2}\left\langle T_1T_2T_3T_4\right\rangle =\mathcal{D}\left(\mathcal{S}F\left(x_{1},x_{2}\right)\right)+\mathcal{S}_{1}^{2}F\left(x_{1},x_{1}\right)+\mathcal{S}_{2}^{2}F\left(x_{2},x_{2}\right)\ .
\]
Here we absorb the two $SO(5)$ indices of $T_{IJ}$ at space-time point $\vec x_i$ using 5d coordinates, $Y_i^I$, thus defining $T_i:=T_{IJ}(\vec x_i)Y_i^IY_i^J$. Then we have defined the following (with  $\vec x_{ij}^2:=(\vec x_{i}-\vec x_{j})^2$ )
\begin{align}
u= x_{1}x_{2}&=\frac{\vec x_{12}^2\vec x_{34}^2}{\vec x_{13}^2\vec x_{24}^2},\,\,\, &v=\left(1-x_{1}\right)\left(1-x_{2}\right)&=\frac{\vec x_{14}^2\vec x_{23}^2}{\vec x_{13}^2\vec x_{24}^2},\,\,\,\notag\\
 y_{1}y_{2}&=\frac{ Y_1.Y_2Y_{3}.Y_{4}}{Y_{1}.Y_{3}Y_{2}.Y_{4}},\,\,\, &\left(1-y_{1}\right)\left(1-y_{2}\right)&=\frac{ Y_1.Y_4Y_{2}.Y_{3}}{Y_{1}.Y_{3}Y_{2}.Y_{4}},\,\,\,\\
g_{ij} &= \frac{Y_i.Y_j}{\vec x_{ij}^2},\,\,\, &\lambda&=x_{1}-x_{2},
\end{align}
and finally
$\mathcal{D}=-\left(\partial_{1}-\partial_{2}+\lambda\partial_{1}\partial_{2}\right)\lambda$, $\mathcal{S}_{i}=\left(x_{i}-y_{1}\right)\left(x_{i}-y_{2}\right)$, and
$\mathcal{S}=\mathcal{S}_{1}\mathcal{S}_{2}$. For more details see \cite{Heslop:2004du}.

In addition to superconformal symmetry, the prepotential $F$ must satisfy the following crossing symmetry constraints: 
\begin{equation}
F(u,v)=F(v,u),\,\,\, F(u/v,1/v)=v^{2}F(u,v).
\label{crossing}
\end{equation}
Note that exchanging $u$ and $v$ corresponds to taking $(x_1,x_2)\rightarrow (1-x_1,1-x_2)$, and taking $(u,v)\rightarrow(u/v,1/v)$ corresponds to taking $\left(x_{1},x_{2}\right)\rightarrow\left(\frac{x_{1}}{x_{1}-1},\frac{x_{2}}{x_{2}-1}\right)$. It is convenient to decompose the the prepotential as follows:
\begin{equation}
F\left(x_{1},x_{2}\right)=\frac{A}{u^{2}}+\frac{g\left(x_{1}\right)-g\left(x_{2}\right)}{u\lambda}+\lambda G\left(x_{1},x_{2}\right),
\label{decomp}
\end{equation}
where the functions $g$ and $G$ can be expanded in terms of CPWs as follows:
\begin{equation}
g(x)=\sum_{m=0}^{\infty}B_{m}g_{m}(x),\,\,\, g_{m}(x)=x^{m+1}{}_{2}F_{1}\left(m+2,m+1,2m+4,x\right)
\label{gcpw}
\end{equation}
\begin{equation}\label{CPW}
\lambda^{2}G\left(x_{1},x_{2}\right)=\sum_{n=0}^{\infty}\sum_{m=n}^{\infty}A_{mn}G_{mn}(x_1,x_2),
\end{equation} 
where the blocks are defined via an expansion as
\begin{equation}
G_{mn}(x_1,x_2)=\sum_{a=0}^{\infty}\sum_{b=0}^{\infty}c_{m+4,n+4}(a,b)\, t_{m+a,n+b}\left(x_{1},x_{2}\right)\ .
\label{Gcpw}
\end{equation}
Here the $t$ functions are Jack polynomials
\[
t_{m,n}\left(x_{1},x_{2}\right)=\left\{ \begin{array}{ll}
-\frac{1}{(m+3)(n+1)}\left(\left(m{-}n{+}1\right)x_{1}^{m+2}x_{2}^{n-1}-\left(m{-}n{+}3\right)x_{1}^{m+1}x_{2}^{n}-x_{1}{\leftrightarrow}x_{2}\right),&\,\,\, n\neq0\\
x_{1}^{m+1}-x_{2}^{m+1},&\,\,\, n=0
\end{array}\right.
\]
and the $c$'s are the coefficients of the blocks expanded in $t$'s and are given explicitly as
\[
c_{m,n+2}(a,b)=\frac{m{-}n{-}1}{\mu{-}1}\left(1-\frac{2a}{\left(m{-}n{-}1\right)\left(\mu{+}1\right)}-\frac{2ab}{\left(m{-}n{-}1\right)\left(\mu{+}1\right)\left(m{+}n\right)}\right)\frac{\left(m_{a}\right)^{2}}{a!\left(2m\right)_{a}}\frac{\left(n_{b}\right)^{2}}{b!\left(2n\right)_{b}}
\]
where
$\mu=m{+}a{-}n{-}b$ and $x_{n}=\Gamma\left(x{+}n\right)/\Gamma(x)$. The indices $m,n$ label operators with leading scaling dimension $m+n$ and spin $m-n$.

The coefficients $B_m$ and $A_{mn}$ in \eqref{gcpw} and \eqref{CPW} are (sums of) squares of OPE coefficients of operators occurring in the $TT$ OPE. The corresponding operators have lowest weight states with $SO(5)$ reps, dimension and spin given as:
	\begin{align}
	\begin{array}{|lll|}\hline
	\mbox{OPE coefficient }  & \mbox{Operator  } \phi_{\Delta,l}^{\text{\bf{rep}}} & \mbox{Example SUGRA operator } \\[-5pt] 
	& &\\
	\hline B_{0} &\phi^{{\bf 14}}_{4,0}& T\\[5pt]
	B_{l+2} &\phi^{{\bf 1}}_{l+4,l}& \text{none}\\[5pt]
	A_{0 0} &\phi^{{\bf 55}}_{8,0}& [TT]_{\bf 55} \ \ \ \  \text{ (half BPS)}\\[5pt]
	A_{l+2\, 0} &\phi^{{\bf 14}}_{l+8,l}& [T\partial^l T]_{\bf 14} \ \ \text{ (protected)}\\[5pt]
	A_{l+2\, 1} &\phi^{\bf 10}_{l+8,l}& [T\partial^l T]_{\bf 10} \ \ \text{ (protected)}\\[5pt]
	A_{n+l\, n\geq2} &\phi^{{\bf 1}}_{l+2n+4,l}& [T\partial^l \Box^{n-2} T]_{\bf 1} \text{ (unprotected)}\\
	\hline
	\end{array}\label{table}
\end{align}
We also present in the last column schematic examples of operators
in each rep which we expect to be present in the supergravity limit according to AdS/CFT. In particular the only operators we expect to develop anomalous dimensions are in the singlet rep and have twist (dimension minus spin) 8 or higher.  Singlet states with twist 4 which appear in the free theory should be absent in the supergravity limit according to the AdS/CFT correspondence (as there are no supergravity states corresponding to these minimum twist reps).

Note that above we have written the blocks $G_{mn}(x_1,x_2)$ as a series expansion. This expansion can be summed up and written explicitly in terms of hypergeometric functions~\cite{Dolan:2003hv}.

\section{Tree-level Supergravity Prediction} \label{sugra}

In the large-$N$ limit, the four-point function can be computed from tree-level Witten  diagrams for supergravity in an $AdS_7 \times S^4$ background. The contributions can be written as the sum of a contribution from free theory (including $N^0$ and $N^{-3}$) and the remaining supergravity contribution (at $N^{-3}$)~\cite{Arutyunov:2002ff}. The corresponding prepotential terms we denote $F^{\rm{free}}$ and $F^{\rm{sugra}}$, respectively. 

\subsection{Free theory Contribution}
Let us first analyze the free contribution: 
\[
F^{\rm{free}}=1+\frac{1}{u^{2}}+\frac{1}{v^{2}}+\frac{1}{N^{3}}\left(\frac{1}{u}+\frac{1}{v}+\frac{1}{uv}\right)
\]
It is not difficult to check that it satisfies \eqref{crossing}. Moreover, decomposing it according to \eqref{decomp} gives 
\begin{equation}
A=1,\,\,\, g(x)=\frac{x}{N^{3}}\left(1+\frac{1}{1-x}\right).
\label{freeg}
\end{equation}
The OPE coefficients obtained from the expansion~\eqref{CPW} are then
\[\textstyle
A^{\rm{free}}_{mn}=
\frac{\left(m+3\right)!\left(m+4\right)!\left(n+1\right)!\left(n+2\right)!\left(m-n+2\right)(m+n+5)}{\left(2m+5\right)!\left(2n+1\right)!}\left(\frac{m+n+6}{72}+\frac{1}{2N^{3}}\frac{\left(-1\right)^{n}}{\left(m-n+1\right)\left(m-n+3\right)\left(m+n+4\right)}\right),
\]
for $m-n$ even and 0 otherwise.

\subsection{Dynamical Contribution}

The prepotential for the 4-point function derived from connected AdS Feynman diagrams is given by~\cite{Arutyunov:2002ff}
\begin{equation}
F^{\rm{sugra}}=-\frac{1}{N^{3}}\frac{\lambda^{2}}{uv}\bar{D}_{3337}.
\label{Fsugrd}
\end{equation}
The $\bar{D}$ functions arise from tree-level Feynman diagrams in AdS and are defined in terms of derivatives of 1-loop massive box functions. For more details, see for example \cite{Arutyunov:2002fh}. It is not difficult to check that \eqref{Fsugrd} satisfies \eqref{crossing} using the following identities:
\[
\bar{D}_{\Delta_{1}\Delta_{2}\Delta_{3}\Delta_{4}}(u,v)=\bar{D}_{\Delta_{3}\Delta_{2}\Delta_{1}\Delta_{4}}(v,u)
\]
\[
\bar{D}_{\Delta_{1}\Delta_{2}\Delta_{3}\Delta_{4}}(u,v)=v^{\Delta_{4}-\Sigma}\bar{D}_{\Delta_{2}\Delta_{1}\Delta_{3}\Delta_{4}}(u/v,1/v),
\]
where $\Sigma=\frac{1}{2}\left(\Delta_{1}+\Delta_{2}+\Delta_{3}+\Delta_{4}\right)$. For a more complete list of identities, see \cite{Arutyunov:2002fh}. Decomposing the supergravity prepotential according to \eqref{decomp} gives 
$A=0$ and
\[
g(x)=\frac{1}{N^{3}}\left(2x_{\,2}F_{1}(2,1,4,x)-\left(x+\frac{x}{1-x}\right)\right).
\]
From \eqref{gcpw} we see that the first term corresponds to the CPW for the stress tensor supermultiplet, and the second term cancels
the free contribution in \eqref{freeg}. This agrees with the discussion around~\eqref{table}  that the only twist 4 operator remaining in the supergravity spectrum is the stress-energy multiplet.

It is convenient to further decompose the function $G$ arising from the decomposition in \eqref{decomp} as follows: 
\begin{align}\label{gsugra}
G^{\text{sugra}}\left(x_{1},x_{2}\right)=G_{{\rm no-log}}\left(x_{1},x_{2}\right)+\log u\, G_{{\rm log}}\left(x_{1},x_{2}\right).
\end{align}
We now wish to perform a CPW expansion at next order in $N^{-3}$. Note that the coefficients  $A_{mn}$ depend on $N$, and the dimension $m+n$ becomes anomalous  and depends on $N$ through the anomalous dimension (with the spin $m-n$ fixed).
Expanding~\eqref{CPW} to first order in $N^{-3}$ we obtain
\begin{equation}\label{CPW2}
\lambda^{2}G^{\text{sugra}}\left(x_{1},x_{2}\right)=\sum_{m=0}^{\infty}\sum_{m=n}^{\infty}\left(A_{mn} \gamma^{\text{sugra}}_{mn} \frac{\partial_m +\partial_n}2 G_{mn}(x_1,x_2)+A^{\text{sugra}}_{mn} G_{mn}(x_1,x_2)\right)\ ,
\end{equation}
where $\gamma^{\text{sugra}}_{mn}$ is the anomalous dimension.%
\footnote{In general there will be more than one operator with the same naive dimension and so here and throughout this paper $\gamma_{mn}$ means an averaged anomalous dimension:  $\gamma^{\text{sugra}}_{mn}= \frac{\sum_i A_{mn;i}\gamma_{mn;i} }{\sum_i A_{mn;i}} $.}
The derivative of the block~\eqref{Gcpw} has the form 
\begin{align}
	(\partial_m +\partial_n)G_{mn}(x_1,x_2) = \log u\, G_{mn}(x_1,x_2)+ \hat G(x_1,x_2),
\end{align}
where $\hat G(x_1,x_2)$ is analytic as $u\rightarrow 0$. 

On equating~\eqref{gsugra} with~\eqref{CPW2}, we see that the coefficients in the CPW expansion of $G_{{\rm log}}$ correspond to $A_{mn}\gamma^{\text{sugra}}_{mn}/2$. Using \eqref{CPW}, one then obtains the following anomalous dimensions:
\[
\textstyle
\gamma_{mn}^{{\rm sugra}}/2=
-\frac{3}{N^{3}}\left(1+\frac{(n-2)(n+1)}{2(m+n+4)(m-n+3)}\right)\frac{(n-1)_{6}}{(m-n+1)(m-n+2)(m+n+5)(m+n+6)},
\]
for spin $ m-n$ even and 
0 for  $m-n$ odd.
Note that they scale like $n^5$ in the large-$n$ limit (with $m-n$, the spin, fixed).

On equating the ``no-log" part of~\eqref{gsugra} with that of~\eqref{CPW2}, we determine the $N^{-3}$ corrections to the OPE coefficients. We verify that they satisfy the relation 
\begin{align}\label{derivrel}
A_{mn}^{{\rm sugra}}=\frac{1}{2}\left(\frac{\partial}{\partial n}+\frac{\partial}{\partial m} \right)\left(A_{mn}^{{\rm free}}\gamma_{mn}^{{\rm sugra}}\right)
\end{align}
seen in other contexts~\cite{Heemskerk:2009pn,Fitzpatrick:2011dm, Alday:2014tsa}. Note in particular that the anomalous dimensions vanish for $n=0,1$, but their derivatives do not. Nevertheless~\eqref{derivrel} still holds! This corresponds to the fact that the  operators with $n=0,1$ have no anomalous dimensions (see~\eqref{table}), but  have normalisations which do depend on $N$ differently to the free theory. This is in contrast to $\cN=4$ SYM where the OPE coefficients of protected operators are also protected and therefore given by the free theory result. This difference will be crucial when deriving corrections to supergravity in the next section.

Note that we find a similar formula to~\eqref{derivrel} for the corrections to the supergravity result we find in the next section.

\section{Corrections to Tree-level Supergravity} \label{corrections}
\subsection{General Considerations}
In this section, we will deduce corrections to the tree-level supergravity prediction for the 4-point correlator of stress-tensor multiplets in the 6d $(2,0)$ theory. We will follow the strategy set out in \cite{Heemskerk:2009pn}, which solved the crossing equations for a 4-point function in a generic 2d or 4d CFT with a large-$N$ expansion to first nontrivial order in $1/N$ by truncating the
spin of the CPW expansion, and showed that the number of solutions
with spin at most $L$ is $(L+2)(L+4)/8$. That paper also provided a simple holographic
explanation for this counting by considering a massive scalar field in AdS with local
quartic interactions (which can be thought of as a toy model for the low-energy effective theory of the gravitational dual), and observing
that up to integration by parts and equations of motion, the local
bulk interactions are in one-to-one correspondence with solutions to
the crossing equations. In particular they showed that there are $L/2+1$ independent quartic
interactions which can create or annihilate a state of spin $L$, with the number of derivatives ranging from $2 L$ to $3L$
(note that only even spins are allowed). For example, there is one
spin-0 interaction vertex $\phi^{4}$, and two spin-2 interaction
vertices $\phi^{2}\left(\nabla_{\mu}\nabla_{\nu}\phi\right)^{2}$
and $\phi^{2}\left(\nabla_{\mu}\nabla_{\nu}\nabla_{\rho}\phi\right)^{2}$
which contain four and six derivatives, respectively. More generally, they argued that a basis of 4-point spin-$L$ interaction vertices  can mapped to a basis of 4-point S-matrices in flat space whose elements take the form
\[
(st)^{L/2}u^{c},\,\,\, c\in\{0,...,L/2\},
\]
where $s,t,u$ are Mandelstam variables satisfying $s+t+u=4m^{2}$.
This corresponds to interaction vertices of the form
\[
\nabla_{\mu_{1}...\mu_{L/2}\nu_{1}...\nu_{L/2}\rho_{1}...\rho_{c}}\phi\nabla_{\mu_{1}...\mu_{L/2}}\phi\nabla_{\rho_{1}...\rho_{c}}\phi\nabla_{\nu_{1}...\nu_{L/2}}\phi,
\]
where we associate $s,t,u$ with $\mu,\nu,\rho$ derivatives, respectively. Hence, the total
number of interactions with spin at most $L$ is $\sum_{l=0}^{L/2}(L/2+1)=(L+2)(L+4)/8$. The authors of \cite{Heemskerk:2009pn} also argued that the large-twist behaviour of the anomalous dimensions is directly related to the number of derivatives appearing in the bulk interactions. This makes it possible to deduce how the solutions to the crossing equations should scale in the large-$N$ limit by analyzing the large-$n$ behaviour of their anomalous dimensions, where $n$ is the twist.

This approach was implemented in $\mathcal{N}=4$ SYM in~\cite{Alday:2014tsa} where
it was shown that in the large-$n$ limit  
\begin{equation}
\gamma^{{\rm spin-0}}/\gamma^{{\rm sugra}}\sim n^{6},\label{eq:n6}
\end{equation}
which suggests that the first correction to 10d supergravity in the low-energy effective action has six more derivatives than the supergravity Lagrangian (stringy corrections to the stress tensor four-point correlator in $\mathcal{N}=4$ SYM were also obtained in \cite{Goncalves:2014ffa}). By dimensional analysis, this must be suppressed
by $(l_p^{10d})^6$ times some function of the string coupling $g_s$, where $l_p^{10d}$ is the 10d Planck length. For example, in tree-level superstring theory this would correspond to an $\alpha'{}^{3}$ correction and at 1-loop it would have the form $G^{10d}_{N}/\alpha'$, where $\alpha' \sim g_s^{-1/2} (l_p^{10d})^2$  is the square of the string length and $G^{10d}_{N} \sim (l_p^{10d})^8$ is Newton's constant in 10d.  In the supergravity approximation, the former would vanish and the latter would correspond to a quadratically divergent 1-loop counterterm. To simplify the discussion, let us fix the value of the string coupling. The contribution to the tree-level 4-point amplitude arising from such an interaction vertex in the low-energy effective action then goes like $G_{N}^{10d} (l_p^{10d})^6 \sim (l_p^{10d})^{14}$. Recalling that $l_p^{10d} \sim N^{-1/4}$ for IIB string theory on $AdS_{5}\times S^{5}$ with $N$ units of flux through the 5-sphere (in units of the AdS radius), we conclude that the spin-0 solution to the crossing equations is suppressed by $N^{-7/2}$. 

Using similar arguments, we can deduce that higher-spin solutions are suppressed by an additional factor of at least $(l_p^{10d})^4 \sim N^{-1}$ compared to the spin-0 solution since they are dual to interaction vertices with at least four more derivatives. Recalling that the spin-0 contribution scales like $N^{-7/2}$ (holding
$g_s$ fixed), we see that higher-spin contributions are suppressed
by at least a factor of $N^{-9/2}$ and are therefore subleading with respect to the 1-loop supergravity correction (recently found in~\cite{Aprile:2017bgs}), which scales like $(G_{N}^{10d})^{2}\sim N^{-4}$.

Since the counting of bulk interaction vertices described in \cite{Heemskerk:2009pn} is not tied to any particular
dimension of AdS, this suggests that the solution counting for the crossing equations
in 2d and 4d should also hold in 6d, which is consistent with our findings below. In particular, we will find that \eqref{eq:n6} is once again satisfied, indicating
that the leading correction to 11d supergravity contains six more
derivatives than the supergravity Lagrangian and must be suppressed by $(l_p^{11d})^{6}$. Hence, the contribution to the 4-point amplitude arising
from this interaction vertex goes like $G_{N}^{11d}(l_p^{11d})^{6} \sim (l_p^{11d})^{15} \sim N^{-5}$,
where we noted that for M-theory in $AdS_{7}\times S^{4}$ with $N$
units of flux through the 4-sphere, $ l_p^{11d} \sim N^{-1/3}$ in units of the AdS radius. Hence, the spin-0 solution to the crossing equations must be suppressed by a factor of $N^{-5}$. Note that for
the 6d $(2,0)$ theory, the only tunable parameter is $N$. Using similar arguments one can deduce that
if the anomalous dimension for a solution to the 6d crossing equation scales like $n^{\alpha}$, then it must be suppressed by
a factor of 
\begin{equation}
G_{N}^{11d}(l_p^{11d})^{\alpha-5}\sim N^{-\left(3+\left(\alpha-5\right)/3\right)}.
\label{scaling}
\end{equation}
From this equation, we see that solutions with spin greater than zero are subleading with respect
to the 1-loop supergravity prediction which scales like $(G_{N}^{11d})^{2}\sim N^{-6}$ (note that the 1-loop supergravity amplitude contains a cubic divergence \cite{Fradkin:1982kf}, so when the corresponding counterterm is lifted to M-theory it is expected to scale like $G_{N}^{11d}/(l_p^{11d})^{3} \sim N^{-5}$).  

In the following subsections, we will present explicit solutions to the
crossing equations for spin $L=0,2,4$, confirming the claims
above. In the language of  section \ref{cpw}, we will look for solutions to the crossing equations for which the coefficients of the CPW expansion are zero for
$m-n> L$. Solutions corresponding to a spin-$L$ truncation will be labelled with subscripts running from $2L$ to $3L$ in increments of 2 (corresponding to the number of derivatives in the corresponding bulk interaction vertices, according to the toy model of  \cite{Heemskerk:2009pn}). 

To obtain the solutions we use the following ansatz (similar to the procedure used for $\cN=4$ SYM in~\cite{Alday:2014tsa}) for the prepotential:
\begin{align}\label{ansatz}
	F(u,v)=\lambda^2 u^a v^b \bar D_{p_1p_2p_3p_4} + \text{ crossing}\ ,
\end{align}
where ``+ crossing" means we sum all 6 terms obtained by permuting the external points. 
We split this into a piece proportional to $\log u$ and a piece analytic as $u\rightarrow 0$ 
\begin{align}
F\left(x_{1},x_{2}\right)=F_{{\rm no-log}}\left(u,v\right)+\log u\, F_{{\rm log}}\left(u,v\right)
\end{align}
and we impose the following small $u$ behaviour: 
\begin{align}\label{constraint}
	F_{{\rm log}}(u,v)= O(u) \qquad 	F_{{\rm no-log}}(u,v)= O(u^0)\ .
\end{align} 
Note that as $u\rightarrow 0$ the blocks~\eqref{Gcpw} are of the form  $G_{mn} = O(u^{n-1})$ for $n>0$.
Thus the first constraint in~\eqref{constraint} arises from insisting that the lowest twist  operators which can develop  anomalous dimensions (controlled by the $\log u$ piece) are the  two-particle supergravity states $T \partial^l T$ in the singlet rep (see~\eqref{table}) with corresponding blocks $G_{mn}$ with $n\geq2$ of $O(u)$. 
However as seen already in the supergravity approximation, the OPE coefficients (controlled by the non-log piece) of protected operators (with $n=0,1$) can be $N$-dependent even though they have no anomalous dimension. These are operators of the form  $T \partial^l T$ but in non-singlet reps and so correspondingly the non-log part of $F$ is $O(u^0)$.

The small $u$ behaviour of the $\bar D$ functions themselves is
\begin{align}
	\bar D_{p_1p_2p_3p_4}|_{\log } &= \left\{\begin{array}{ll}
	O(u^p) &p \geq 0\\
	O(u^0) &p < 0
	\end{array}
	\right.
	\\
	\bar D_{p_1p_2p_3p_4}|_{\rm no-log } &= \left\{\begin{array}{ll}
	O(1/u^p) &p \geq 0\\
	O(u^0) &p < 0 \qquad \qquad p:=\frac12 (-p_1-p_2+p_3+p_4).
	\end{array}
	\right.
\end{align}
Using this we find that the constraint~\eqref{constraint} (including all crossing channels) yields six inequalities on the 6 parameters $a,b,p_i$ of~\eqref{ansatz}:
\begin{dmath}
\left\{2 a\geq \max \left(0,p_1+p_2-p_3-p_4\right),2 b\geq \max \left(0,-p_1+p_2+p_3-p_4\right),-2 a-2
b+p_1+p_2+p_3-p_4-8\geq \max \left(0,p_1-p_2+p_3-p_4\right),\max \left(0,-p_1-p_2+p_3+p_4\right)+2 a\geq 2,\max
\left(0,p_1-p_2-p_3+p_4\right)+2 b\geq 2,\max \left(0,-p_1+p_2-p_3+p_4\right)-2 a-2 b+p_1+p_2+p_3-p_4-8\geq 2\right\}\, .
\end{dmath}
The sum of these six inequalities yields a simple constraint on the sum of the indices of the $\bar D$ function
\begin{align}\label{bound}
	p_1+p_2+p_3+p_4 \geq 22\ .
\end{align}
The $\bar D$ functions satisfy a number of identities, which can be used to reduce any $\bar D$ function into one of the forms $\bar  D_{nnnn}, \bar D_{n+1\,n+1\,nn}, \bar D_{nnn\,n+2}, \bar D_{nn\,n+1\,n-1}$~\cite{Arutyunov:2002fh}. 
Inputting  these forms into  the above inequalities allows us to enumerate all possible solutions, although there are still further identities satisfied by the solutions obtained via the above procedure which we manually impose to reduce the solution space further. We then order the solutions thus derived by their value of $\sum_i p_i$.

In the next subsections, we give the first few cases in this list and perform a CPW expansion of them. Very nicely, we find that the CPW expansions do indeed all truncate to small spin, the first to spin-0, the next two to spin-2 and the next three to spin-4 etc. We further find that in all cases the corrections to the OPE coefficients are related to the anomalous dimensions by the analogous  formula to the supergravity case~\eqref{derivrel}
\begin{align}\label{derivrel2}
A_{mn}^{{\rm corr.}}=\frac{1}{2}\left(\frac{\partial}{\partial m}+\frac{\partial}{\partial n}\right)\left(A_{mn}^{{\rm free}}\gamma_{mn}^{{\rm corr.}}\right)\ .
\end{align}

\subsection{Spin-Zero Truncation}
The solution with smallest value of $\sum p_i =22$ is 
\[
F^{\rm{spin-0}}=\lambda^{2}uv\bar{D}_{5755}\ .
\]
Note that the overall numerical coefficient is unfixed. Performing a CPW expansion of this we find that it indeed corresponds to the spin-0 truncation of the 6d crossing equations. Indeed, using \eqref{decomp} and \eqref{CPW}, we deduce the following (averaged) anomalous dimensions:
\[
\gamma_{n,n}^{{\rm spin-0}}=-\frac{(n-1)_8 \,n_{6} }{2240(2n+3)(2n+5)(2n+7)}.
\]
Note that they scale like $n^{11}$ in the large-$n$ limit. Using \eqref{scaling}, we see that this solution should therefore be suppressed by a factor of $N^{-5}$. 

\subsection{Higher-Spin Truncations}

The next two solutions are 
\[
F_{4}^{{\rm spin-2}}=2\lambda^{2}uv\left(\bar{D}_{6776}+\bar{D}_{7676}+\bar{D}_{7766}\right),\,\,\, F_{6}^{{\rm spin-2}}=6\lambda^{2}uv\bar{D}_{7777},
\]
where the overall numerical coefficients are once again unfixed.
We find that these two solutions correspond to spin-2 truncations of the 6d crossing equations.  
 The anomalous dimensions for the first solution $F_{4}^{{\rm spin-2}}$ are given by
\[
\frac12\gamma_{(4)\, n+2,n}^{{\rm spin-2}}=\frac{(n-1)_{10}\,n_{8}}{1935360(2n+3)(2n+5)(2n+7)(2n+9)(2n+11)}
\]
\[
\frac12\gamma_{(4)\, n,n}^{{\rm spin-2}}=\frac{(n-1)_8\,n_{6}\left(19n^{6}+285n^{4}+4665n^{3}-3509n^{2}-46170n-21744\right)}{483840(2n+1)(2n+3)(2n+5)(2n+7)(2n+9)}.
\]
Moreover the anomalous dimensions for the second solution $F_{6}^{{\rm spin-2}}$ are
\[
\frac12\gamma_{(6)\, n+2,n}^{{\rm spin-2}}=-\frac{\left((n-1)_{10}\right)^{2}(n+3)_2}{6451200(2n+3)(2n+5)(2n+7)(2n+9)(2n+11))}
\]
\[
\frac12\gamma_{(6)\, n,n}^{{\rm spin-2}}=\frac{\left((n-1)_{8}\right)^{2}\left(5n^{6}+75n^{5}+877n^{4}+5645n^{3}+20754n^{2}+41020n+16032\right)}{1612800(2n+1)(2n+3)(2n+5)(2n+7)(2n+9)}
\]
Note that the anomalous dimensions for $F_{4}^{{\rm spin-2}}$  and $F_{6}^{{\rm spin-2}}$ scale like $n^{15}$ and $n^{17}$, respectively. This agrees with the holographic arguments of \cite{Heemskerk:2009pn}, which predict two spin-2 solutions which scale like $n^4$ and $n^6$ times the spin-0 solution since the corresponding bulk interaction vertices contain four and six additional derivatives, respectively. Plugging these large-$n$ scalings into \eqref{scaling}, we then find that $F_{4}^{{\rm spin-2}}$  and $F_{6}^{{\rm spin-2}}$ should be suppressed by factors of $N^{-19/3}$ and $N^{-7}$, respectively, and are therefore subleading with respect to 1-loop supergravity correction which should scale like $N^{-6}$ but has not been computed yet. 	
	
Finally, the next three solutions are
	\begin{align}
	F_{8}^{{\rm spin-4}}&=
	u v \left(u \bar{D}_{7887}+u \bar{D}_{8787}+v \bar{D}_{8787}+v
	\bar{D}_{8877}+\bar{D}_{7887}+\bar{D}_{8877}\right)\notag\\
	F_{10}^{{\rm spin-4}}&=2 u v
	\bar{D}_{8888} (u+v+1)\notag\\
	F_{12}^{{\rm spin-4}}&=2 u v \left(u^2 \bar{D}_{9988}+v^2
	\bar{D}_{8998}+\bar{D}_{9898}\right).
	\end{align}
As the labelling suggests, one finds that all three of these truncate at spin-4 in a CPW expansion. The averaged anomalous dimensions computed from the first one are
\begin{align}
	\frac12\gamma_{(8)\,n+4,n}^{{\rm spin-4}} &= \frac{(n-1)_{12} n_{10}(n+3)_4 }{2128896000 (2 n+3) (2 n+5) (2 n+7) (2 n+9) (2 n+11) (2 n+13) (2
		n+15)}\\
	\frac12\gamma_{(8)\,n+2,n}^{{\rm spin-4}} &=\frac{(n-1)_{10} n_8 (n+3)_2  \,p_6(n)}{354816000 (2
		n+1) (2 n+3) (2 n+5) (2 n+7) (2 n+9) (2 n+11) (2 n+13)}\\
	\frac12\gamma_{(8)\,n,n}^{{\rm spin-4}}&=\frac{(n-1)_8 n_6  \,p_{12}(n)}{141926400 (2 n-1) (2 n+1) (2 n+3) (2 n+5) (2 n+7) (2 n+9) (2 n+11)},
\end{align}
where
\begin{dmath}
	p_6(n)=\left(11 n^6+231 n^5+2081 n^4+10269 n^3+20880 n^2-12992 n-10580\right)
	\end{dmath}
\begin{dmath}
	p_{12}(n)=\left(161 n^{12}+4830 n^{11}+62823 n^{10}+463700 n^9+1924951 n^8+2150270 n^7-21226599 n^6-119491360
	n^5-323505448 n^4-802014480 n^3-1360491344 n^2+209249280 n+360009216\right)\ .
\end{dmath}
The averaged anomalous dimensions of the second spin-4 truncated function are
\begin{align}
\frac12\gamma_{(10)\,n+4,n}^{{\rm spin-4}} &= -\frac{((n-1)_{12})^2 (n+3)_4 }{12773376000 (2 n+3) (2 n+5) (2 n+7) (2 n+9) (2 n+11) (2 n+13)
	(2 n+15)}\\
\frac12\gamma_{(10)\,n+2,n}^{{\rm spin-4}} &=-\frac{((n-1)_{10})^2 (n+3)_2 \,q_6(n)}{6386688000
	(2 n+1) (2 n+3) (2 n+5) (2 n+7) (2 n+9) (2 n+11) (2 n+13)}\\
\frac12\gamma_{(10)\,n,n}^{{\rm spin-4}}&=\frac{((n-1)_8)^2 q_{12}(n)}{2554675200 (2 n-1) (2 n+1) (2 n+3) (2 n+5) (2 n+7) (2 n+9) (2
	n+11)},
\end{align}
where
\begin{dmath}
	q_6(n)=\left(13 n^6+273 n^5+1543 n^4-693 n^3-12380 n^2+39564 n+22530\right)
	\end{dmath}
\begin{dmath}
	q_{12}(n)=\left(125 n^{12}+3750 n^{11}+66609 n^{10}+805850 n^9+6848571 n^8+41875170 n^7+194627827
	n^6+726165030 n^5+2149177860 n^4+4642708600 n^3+5264600928 n^2-1200615360 n-1456584192\right).
\end{dmath}
The averaged anomalous dimensions of the third spin-4 truncated function are
\begin{align}
\frac12\gamma_{(12)\,n+4,n}^{{\rm spin-4}} &=-\frac{((n-1)_{12})^2 (n+3)_4  \left(3 n^2+27 n+233\right)}{332107776000 (2 n+3) (2 n+5) (2
	n+7) (2 n+9) (2 n+11) (2 n+13) (2 n+15)}\\
\frac12\gamma_{(12)\,n+2,n}^{{\rm spin-4}} &=\frac{((n-1)_{10})^2  (n+3)_2  \,r_8(n)}{166053888000 (2 n+1) (2 n+3) (2 n+5) (2 n+7) (2 n+9) (2 n+11) (2 n+13)}\\
\frac12\gamma_{(12)\,n,n}^{{\rm spin-4}}&=-\frac{((n-1)_8)^2  r_{14}(n)}{66421555200 (2 n-1) (2
	n+1) (2 n+3) (2 n+5) (2 n+7) (2 n+9) (2 n+11)},
\end{align}
where
\begin{dmath}	r_8(n)=\left(21 n^8+588 n^7+6631 n^6+38409 n^5+282976 n^4+2472015 n^3+10181116
	n^2+14543844 n+5050800\right)
	\end{dmath}
\begin{dmath}
	r_{14}(n)=\left(105 n^{14}+3675 n^{13}+51200 n^{12}+341625 n^{11}-1956606 n^{10}-72462025 n^9-829217484
	n^8-5712467805 n^7-26894812571 n^6-90925907190 n^5-224780417604 n^4-403755466040 n^3-405122118336 n^2+106577896320 n+115817748480\right).
	\end{dmath}

We see that the anomalous dimensions for $F_{8}^{{\rm spin-4}}$,$F_{10}^{{\rm spin-4}}$   and $F_{12}^{{\rm spin-4}}$ scale like $n^{19}$, $n^{21}$ and $n^{23}$ respectively. This also agrees with the holographic arguments of \cite{Heemskerk:2009pn}, which predict three spin-4 solutions which scale like $n^8$, $n^{10}$, $n^{12}$ times the spin-0 solution since the corresponding bulk interaction vertices contain 8, 10 and 12 additional derivatives, respectively. Plugging these large-$n$ scalings into \eqref{scaling}, we then find that $F_{8}^{{\rm spin-4}}$, $F_{10}^{{\rm spin-4}}$, $F_{12}^{{\rm spin-4}}$ should be suppressed by factors of $N^{-23/3}$, $N^{-25/3}$, $N^{-9}$ respectively.

\section{Conclusion} \label{conclusion}
 
We have explored some implications of superconformal and crossing symmetry for 4-point correlators of stress tensor multiplets in the 6d $(2,0)$ theory in order to gain new insight into M-theory beyond the supergravity approximation. We did so by finding crossing symmetric functions with single discontinuities which have CPW expansions that truncate to finite spin. Adapting the holographic arguments developed in \cite{Heemskerk:2009pn}, we deduced how these solutions scale in the the large-$N$ limit by studying the large-twist behaviour of their anomalous dimensions and find that while the spin-0 solution scales like $N^{-5}$, all higher-spin solutions are suppressed by at least $N^{-19/3}$. As a result, the spin-0 solution corresponds to the leading correction to the supergravity prediction for the 4-point correlator (which scales like $N^{-3}$) and all higher-spin solutions are subleading compared to the 1-loop supergravity prediction (which scales like $N^{-6}$ but has not been computed yet). 

Our results provide important hints about the low energy effective action of M-theory on $AdS_7 \times S^4$. In particular, noting that the anomalous dimensions of the spin-0 solution scale like $n^6$ compared to those of the supergravity prediction (where $n$ is the twist), this suggests that the corresponding terms in the effective action contain six more derivatives than the supergravity Lagrangian, and are subsequently of the form $\mathcal{R}^4$, where $\mathcal{R}$ is the Riemann tensor. M-theoretic corrections to 11d supergravity of this form were previously deduced in flat background using arguments based on dimensional reduction to string theory \cite{Green:1997di,Green:1997as,Russo:1997mk}, so it would be very interesting to lift them to $AdS_7 \times S^4$ and check that when they are added to the Lagrangian for 11d supergravity in this background, the resulting 4-point amplitude agrees with the spin-0 solution we obtained. It would also be interesting to see if the analogous spin-0 solution for $\mathcal{N}=4$ SYM obtained in \cite{Alday:2014tsa} can be derived by adding $\mathcal{R}^4$ terms to IIB supergravity in $AdS_5 \times S^5$.

As mentioned above, after the spin-0 solution to the crossing equations, the next correction to the 4-point correlator corresponds to a 1-loop supergravity amplitude in $AdS_7 \times S^4$. It would therefore be very interesting to compute this correction. This was recently done for  $\mathcal{N}=4$ SYM for the stress-tensor multiplet correlator in~\cite{Aprile:2017bgs} and a correlator involving the next higher charge half-BPS operator in~\cite{Aprile:2017qoy}. This involved disentangling the contributions of degenerate operators to the four point function~\cite{Alday:2017xua,Aprile:2017qoy} by using free and supergravity-corrected correlation functions of the higher charge half BPS operators (corresponding to higher Kaluza-Klein mode supergravity states) which were recently obtained in~\cite{1608.06624,1710.05923}. An alternative approach which directly gives the data contained in the correlators has also recently been developed~\cite{Alday:2016njk,Caron-Huot:2017vep,Alday:2017xua,Alday:2017vkk}.
 
In order to apply these methods to the $(2,0)$ theory, we would thus need to first generalize the tree-level supergravity prediction for the 4-point correlator of stress tensor multiplets to operators with higher R-charges (the first steps in this direction were recently taken in \cite{Rastelli:2017ymc,Zhou:2017zaw}) and also have better control of the CPWs for higher charge correlators. 
Ultimately, we hope that this approach will provide a systematic way to construct correlation functions of the 6d $(2,0)$ theory, or equivalently the low-energy effective action of M-theory on $AdS_7 \times S^4$, from first principles. 

\begin{center}
\textbf{Acknowledgements}
\end{center}

We thank Fernando Alday, Francesco Aprile, Agnese Bissi, James Drummond, Tomasz Lukowski, and Hynek Paul for useful conversations. AL is supported by the Royal Society as a Royal Society University Research Fellowship holder, and PH by an STFC Consolidated
Grant ST/P000371/1. Both authors express appreciation for MIAPP hosting the workshop "Mathematics and Physics of Scattering Amplitudes" where this work was
initiated.

\end{document}